\documentclass[%
 reprint,
 superscriptaddress,
 showpacs,
 amsmath,amssymb,
 aps,
 prd,
]{revtex4-1}

\usepackage{graphicx}
\usepackage{dcolumn}
\usepackage{bm}
\usepackage{hyperref}

\usepackage{simplewick}

\DeclareMathOperator{\disc}{Disc}

\begin{document}


\title{$a_{0}(980)$ revisited}

\author{Thomas \surname{Wolkanowski}}
 \email{wolkanowski@th.physik.uni-frankfurt.de}
\affiliation{%
 Institute for Theoretical Physics, Goethe University, D-60438 Frankfurt am Main, Germany}%


\author{Francesco \surname{Giacosa}}
 \email{fgiacosa@ujk.edu.pl}
\affiliation{%
 Institute for Theoretical Physics, Goethe University, D-60438 Frankfurt am Main, Germany}%
\affiliation{
 Institute of Physics, Jan Kochanowski University, PL-25406 Kielce, Poland}%
 
\author{Dirk H.\ \surname{Rischke}}
 \email{drischke@th.physik.uni-frankfurt.de}
\affiliation{%
 Institute for Theoretical Physics, Goethe University, D-60438 Frankfurt am Main, Germany}%


\date{\today}

\begin{abstract}
Light scalar hadrons can be understood as dynamically generated resonances. These
arise as `companion poles' in the propagators of quark-antiquark seed states 
when accounting for meson-loop contributions to the self-energies of the latter. 
Along this line, we extend previous calculations of 
T\"ornqvist and Roos and of Boglione and Pennington, where the resonance 
$a_{0}(980)$ appears as companion pole in the propagator of $a_{0}(1450)$
which is predominantly a quark-antiquark state. We also construct an effective 
Lagrangian where $a_{0}(1450)$ couples to pseudoscalar mesons with 
both non-derivative and derivative interactions. Computing the one-loop self-energy,
we demonstrate that the propagator has two poles: a companion pole corresponding
to $a_{0}(980)$ and a pole of the seed state $a_{0}(1450)$. The positions of these
poles are in quantitative agreement with experimental data.
\end{abstract}

\maketitle


\section{\label{sec:section1}Introduction}

The majority of mesons can be understood as being predominantly quark-antiquark 
states \cite{olive}. Yet, various unconventional mesonic states such as glueballs, 
hybrids, and four-quark states are expected \cite{amslerrev,*amslerrev2}. One particular type of 
four-quark meson is that of a `dynamically generated' state 
\cite{tornclose,pelaez3,*pelaez,*pelaez2,oller,*oller2,*oller3,oller4,oller5,*oller6,2006beveren,morgan,dullemond,tornqvist,*tornqvist2,boglione,pennington,zhiyong}. 
Although there are recent works on the compositeness of (dynamically
generated) resonances -- see \emph{e.g.} Refs.\ \cite{guo,ollersigma}, where the former also provides a 
method to quantify the weight of the two-body components of a resonance -- a generally accepted definition of 
dynamical generation does not exist \cite{giacosaDynamical}. 
An interesting version of this idea was put forward in Refs.\ 
\cite{dullemond,tornqvist,*tornqvist2,boglione,pennington}. Consider, for instance, 
a single seed state, \emph{e.g.}\ a quark-antiquark meson with certain quantum numbers. 
This state interacts with other mesons, giving rise to loop contributions
in the corresponding self-energy. These shift the pole of the seed state and, moreover,
a(t least one) new pole may appear. The latter, also denoted as a companion pole, 
corresponds to a dynamically generated resonance. As a consequence, two resonances 
have emerged from a single seed state. In this work, we aim to review some 
previous works and deepen the understanding of the resonance $a_{0}(980)$ as a dynamically 
generated state.

There is a growing consensus that the scalar resonances $f_{0}(1370)$, $f_{0}(1500)$, 
$K_{0}^{\ast}(1430)$, and $a_{0}(1450)$
are predominantly quark-antiquark states; see for example Refs.\
\cite{close,*close2,*close3,*close4,*close5,*close6,tqmix,eLSM1,*eLSM1-2,*eLSM1-3,eLSM2,oller4,oller5,*oller6,zhiyong} 
[recent studies \cite{stani,chenlattice} 
agree in interpreting the resonance $f_{0}(1710)$ as predominantly gluonic]. 
Then, the light scalar states $f_{0}(500)$, $f_{0}(980)$, 
$K_{0}^{\ast}(800)$, and $a_{0}(980)$ are (most likely) predominantly four quark-states 
[see \emph{e.g.} Refs.\ 
\cite{jaffe,*jaffe2,maiani,hooft,fariborz,*fariborz2,rodriguez,molecular,*molecular2,*molecular3,*molecular4,*molecular5,oller,*oller2,*oller3,oller4,oller5,*oller6,tqmix,zhiyong} and refs.\ therein]. 
As we shall show in the following for the heavy scalar--isovector seed state 
$a_{0}(1450)$, the coupling of this state to $\pi\eta,K\bar{K}$, and 
$\pi\eta^{\prime}$ dynamically generates the light $a_{0}(980)$ as a particular type of
four-quark meson.

T\"ornqvist and Roos \cite{tornqvist,*tornqvist2} (in the following denoted as TR) and later 
Boglione and Pennington \cite{boglione,pennington} (denoted as BP)
studied the mechanism of dynamical generation through 
meson-loop contributions to the self-energy. Here, we extend their studies (Sec.\
\ref{sec:section2}) and compare numerical results for the poles of the propagator
to the latest experimental data \cite{olive}. It turns out that (depending on the assignment 
of the poles to physical resonances) the widths of both the seed state 
$a_{0}(1450)$ and the dynamically generated state $a_{0}(980)$ are
by a factor of $2$ larger than the experimental values.
Moreover, the mass of $a_{0}(1450)$ is too large (by $100$ MeV in TR and by $400$ MeV 
in BP). It thus seems that, while qualitatively feasible, the dynamical generation of
resonances as companion poles in the propagator does not yield results
that are in quantitative agreement with experimental data.

In this work, we show that this is actually not true and that the mechanism of dynamical 
generation produces results which are in quantitative agreement with the data. 
To this end, we introduce a Lagrangian inspired by the recently developed extended 
Linear Sigma Model (eLSM) \cite{eLSM1,*eLSM1-2,*eLSM1-3,eLSM2}. Here, the mesons interact
via derivative and non-derivative couplings (Sec.\ \ref{sec:section3}). In our case, the
Lagrangian contains a single scalar--isovector $a_{0}$ seed state which corresponds to the 
resonance $a_{0}(1450)$. A careful analysis of the pole structure of the corresponding
propagator shows that it is indeed possible to obtain a narrow resonance with 
mass around $1$ GeV, the 
pole coordinates of which fit quite well with those of the physical $a_{0}(980)$ resonance, 
and \emph{simultaneously} obtain a pole for the seed state in agreement with that
for the $a_{0}(1450)$ \cite{olive}. Finally, we also investigate the use of Feynman rules 
in the context of quantum field theories with derivative interactions, and demonstrate 
that for a particular form of the Lagrangian there may be a discrepancy between ordinary 
Feynman rules and dispersion relations (see App.\ \ref{app:appendix1}).

Our units are $\hbar=c=1$. The metric tensor is $\eta_{\mu \nu} = \text{diag}(+,-,-,-)$.

\section{\label{sec:section2}Dynamical generation}

\subsection{Approach of TR and BP}

Following earlier work \cite{dullemond}, T\"{o}rnqvist \emph{et al.}\ studied the scalar sector 
in a 
unitarized quark model by including meson-loop contributions \cite{tornqvist,*tornqvist2}. They showed
that meson-loop effects may serve to explain the existence of the light scalar states.

The following two points are relevant in the mechanism of dynamical generation,
irrespective of the quantum numbers of the hadronic resonance considered: 
$(i)$ The propagator of a quark-antiquark seed state gets dressed by meson-loop
contribution to the self-energy. These contributions shift the mass of the state and 
change the form of its spectral function. When increasing the coupling, 
the corresponding pole moves away from the real axis and follows a certain trajectory 
in the complex plane. The mass and the width of the resonance are determined by the position 
of the complex pole of the dressed propagator on the appropriate 
Riemann sheet -- a procedure first proposed by Peierls a long time ago \cite{peierls}. 
$(ii)$ If the interaction exceeds a critical value, a(t least one) companion pole can appear 
in the complex plane. If this pole is sufficiently close to the real axis, it can manifest itself
in the spectral function as an \emph{additional} resonance with the same quantum numbers
as the seed state \cite{morgan,tornqvist,*tornqvist2,polosa}. 
Since the coupling of scalars to pseudoscalars is large,
the scalar sector is particularly affected by such distortions of the spectral function. 

We now recapitulate the seminal works TR \cite{tornqvist,*tornqvist2} and BP \cite{pennington},
where the latter uses the same model as the former but with a slightly different set of 
parameters. The main goal is the determination of the inverse propagator of a 
resonance after applying a Dyson resummation of loop contributions to the self-energy:
\begin{equation}
\Delta^{-1}(s)=s-m_{0}^{2}-\Pi(s)\ ,
\end{equation}
where $s$ is the first Mandelstam variable, $m_{0}$ is the
bare mass of the seed state, and $\Pi(s)=\sum_{i}\Pi_{i}(s)$ is the self-energy
\footnote{Note that we use a different sign convention 
for the propagator $\Delta(s)$ and the self-energy $\Pi(s)$ than Refs.\
\cite{tornqvist,*tornqvist2,pennington}. In the first reference the scattering amplitude is studied, 
but only its denominator is important here, which equals the inverse propagator.}.
Here, the sum runs over the loops emerging from the coupling of the resonance
to various mesons. The imaginary part of $\Pi_i(s)$ corresponds to the partial
decay width of the resonance into mesons in channel $i$.
The real part of $\Pi(s)$ on the real axis is related to the imaginary part by the 
dispersion relation
\begin{equation}
\operatorname{Re}\Pi(s)=\frac{1}{\pi} -\hspace{-0.395cm}\int\text{d}s^{\prime} \ 
\frac{-\operatorname{Im}\Pi(s^{\prime})}{s-s^{\prime}} \ .
\label{eq:disp}
\end{equation}
TR and BP now assume a simple model for the imaginary part of $\Pi_i(s)$, 
see Refs.\ \cite{tornqvist,*tornqvist2,boglione,pennington,harada,*harada2} for details:
\begin{equation}
\operatorname{Im}\Pi_i(s) 
=-g_{i}^{2}\frac{k_{i}(s)}{\sqrt{s}}(s-s_{A,i})F_{i}^{2}(s)\Theta(s-s_{th,i}) \ .
\label{eq:ImPi}
\end{equation}
In the scalar--isovector sector the Adler zeros $s_{A,i}$ are set to zero for simplicity 
\cite{tornqvist,*tornqvist2,pennington}. The form factor is chosen to be a simple exponential,
\begin{equation}
F_{i}(s)=\exp[-k_{i}^{2}(s)/(2k_{0}^{2})] \ ,
\end{equation}
where $k_{0}$ is a cutoff parameter, and $k_{i}(s)$ is the absolute value of the 
three-momentum of the decay particles in the rest frame of the resonance,
\begin{equation}
k_{i}(s) = \frac{1}{2\sqrt{s}}\sqrt{s^{2}+(m_{i1}^{2}-m_{i2}^{2})^{2}-2(m_{i1}^{2}+m_{i2}^{2})s} \ .
\end{equation}
Here, $m_{i1},m_{i2}$ are the masses of the decay particles, \emph{i.e.}, in our case
the pseudoscalar mesons
\footnote{TR and BP did not quote values for $m_{i1},\, m_{i2}$. 
In this work, we consistently used the isospin-averaged numerical values 
given in the PDG from 2002, the year of publication of BP. Note that these values 
differ from the ones used in TR, so that our results for the pole positions 
slightly differ numerically from theirs.}. 
The function $F_{i}(s)$ guarantees that the imaginary part of $\Pi(s)$ vanishes 
sufficiently fast for $s \rightarrow \infty$ (the inverse cutoff $k_0$ corresponds 
to the non-vanishing size of a typical hadron). The step function in Eq.\ (\ref{eq:ImPi}) 
ensures that the decay channel $i$ contributes only when the squared energy of the
resonance exceeds the threshold value $s_{th,i}$. Finally, the coupling constants 
$g_{i}$ are related by $SU(3)$--flavor symmetry.

Note that one may also define the so-called Breit--Wigner mass of a resonance
as the real-valued root of the real part of the inverse propagator, $\operatorname{Re}\Delta^{-1}(s)=0$. These roots can be
found by identifying the intersections of the so-called `running mass'
\begin{equation}
m^{2}(s)=m_{0}^{2}+\operatorname{Re}\Pi(s) 
\label{eq:runningmass}
\end{equation}
with the straight line $f(s)=s$,
where $s$ is purely real. This definition of the mass of the resonance is also used in
TR and BP. However, the Breit--Wigner 
mass does not necessarily correspond to a pole in the complex energy plane or to a
peak in the spectral function.

For the scalar--isovector sector, the main results of TR and BP can be summarized
as follows:
\begin{enumerate}
\item TR found a pole on the second Riemann sheet with coordinates  
\footnote{We apply the usual parameterization for propagator poles, 
$s_{\text{pole}}=m_{\text{pole}}^{2}-i\hspace{0.02cm}m_{\text{pole}}\Gamma_{\text{pole}}$ \ .} 
$m_{\text{pole}}=1.084$ GeV and $\Gamma_{\text{pole}}=0.270$ GeV, which is a 
companion pole corresponding to the resonance $a_{0}(980)$. A reanalysis 
[the second paper quoted in Ref.\ \cite{tornqvist,*tornqvist2}] where
the complex plane was investigated more carefully revealed another pole 
with $m_{\text{pole}}=1.566$ GeV and $\Gamma_{\text{pole}}=0.578$ GeV on the third 
sheet. This pole is indeed the original seed state and describes the resonance 
$a_{0}(1450)$. It was suggested that, although the numerical agreement was not yet 
satisfactory, an improved model could in principle be capable of describing the whole 
scalar--isovector sector up to $1.6$ GeV. TR also reports one (but not more) intersection 
point(s) of the running mass from Eq.\ (\ref{eq:runningmass}).
\item BP used the same approach, but did not look for poles of the propagator.
Instead, they considered the Breit--Wigner mass. Compared to TR, also the 
values of the bare mass parameter $m_{0}$ as well as the overall strength of the couplings
$g_i$ in Eq.\ (\ref{eq:ImPi}) were changed. 
BP found two intersection points for the running mass from Eq.\ 
(\ref{eq:runningmass}), one in the region around $1$ GeV 
corresponding to $a_{0}(980)$ (like TR) and another one at about $1.4$ GeV 
(absent in TR). This latter intersection was interpreted as the state $a_{0}(1450)$.
Note that, although BP did not investigate the poles of the propagator, a pole and an
 intersection were reported in an earlier work \cite{boglione}.
\end{enumerate}
Apparently, the situation is somewhat inconclusive regarding the number and location
of poles of the propagator and/or intersection points of the running mass. 
Therefore, we decided to repeat the study of TR and BP and investigate
the propagator in the complex plane including all Riemann sheets nearest to the
first (physical) sheet in order to clarify this problem.
The self-energy on the unphysical sheet(s) is obtained by analytic continuation.
To this end, one first computes the discontinuity of the self-energy across the
real $s$-axis,
\begin{equation}
\disc\Pi(s)=2i\lim_{\epsilon\rightarrow0^{+}}\sum_{i}
\operatorname{Im}\Pi_{i}(s+i\epsilon) \ , \ \ \ s \in \mathbb{R} \ .
\end{equation}
Then, the appropriately continued self-energy $\Pi^{c}(s)$ on the next Riemann sheet is 
obtained via
\begin{equation}
\Pi^{c}(s)=\Pi(s)+\disc\Pi(s) \ .
\end{equation}
This expression is valid on the whole Riemann sheet, \emph{i.e.}, $s$ is complex-valued.
Note that in our case there are three thresholds, in successive order corresponding 
to the decays of $a_0$ into $\pi\eta$, $K\bar{K}$, and $\pi\eta^{\prime}$. These channels will
be numbered $i=1,2,3$ in the following.
Thus, crossing the real $s$-axis at values of $s$ in the interval $(s_{th,1},s_{th,2}]$, we
move from the first to the second Riemann sheet, in the following denoted by roman
numeral II. Analogously, crossing the real
$s$-axis in the interval $(s_{th,2},s_{th,3}]$, we move from the first to the third (III) sheet.
Finally, crossing the real $s$-axis at $s > s_{th,3}$, we move from the first
to the sixth (VI) sheet (in the standard notation). Since we will also show plots of 
the spectral function $d(x)$, we recall its definition,
\begin{equation}
d(x)=-\frac{2x}{\pi}\lim_{\epsilon\rightarrow0^{+}}\operatorname{Im}
\Delta(x^{2}+i\epsilon) \ , 
\end{equation}
where $x=\sqrt{s}$.

\subsection{Spectral functions and poles}

We introduce a dimensionless parameter 
$\lambda \in [0,1]$ and replace the coupling constants in Eq.\ (\ref{eq:ImPi}) by
$g_{i}^{2}\rightarrow \lambda g_{i}^{2}$. In consequence, for $\lambda =0$ the self-energy
vanishes and we just obtain the spectral function of the non-interacting
seed state, \emph{i.e.}, a delta function. The corresponding pole lies on the real 
$\sqrt{s}$-axis.
Increasing $\lambda$ from zero to $1$, the interaction is successively increased and
we can monitor in a controlled manner how the spectral function changes. In the
following figures, we will show the spectral function for the physical value $\lambda=1.0$ 
and for the intermediate value $\lambda=0.4$.
Changing $\lambda$ from zero to one, we will also see how the pole of the seed state 
moves off the real axis and other poles emerge, which correspond to the 
dynamically generated resonances. A continuous
change of $\lambda$ will trace out pole trajectories in the complex $\sqrt{s}$-plane. 
The final and physical locations of the poles are reached
when $\lambda=1.0$, which we indicate by a dot in the following figures.
We consider the three Riemann sheets nearest to the physical region (\emph{i.e.}, 
the first sheet) 
in one figure (a list of the poles corresponding to the resonances of interest 
can be found in Sec.\ \ref{sec:conc}). 
For TR, we use the values $g_{1}=1.2952$ GeV, $g_{2}=0.8094$ GeV, $g_{3}=0.9461$ GeV, and 
$k_{0}=0.56$ GeV, and for BP $g_{1}=1.7271$ GeV, $g_{2}=1.0975$ GeV, $g_{3}=1.4478$ 
GeV, and $k_{0}=0.56$ GeV.

The results are shown in Fig.\ \ref{fig:spec_and_poles}. We first discuss the results for
the TR parameterization, and then those for BP:
\begin{enumerate}
\item The two panels in the upper row of Fig.\ \ref{fig:spec_and_poles} show the 
results of TR. For $\lambda=1.0$ the spectral function exhibits a narrow peak in the region 
around $1$ GeV that was interpreted by TR as the $a_{0}(980)$ resonance. 
We furthermore observe a broad structure above $1.5$ GeV. For decreasing coupling 
strength the narrow peak around $1$ GeV vanishes, while the broad structure becomes more 
pronounced. It is located around $1.4$ GeV, which is the location of the
seed state. The width of the peak decreases with $\lambda$, such that we obtain
a delta function for $\lambda =0$, as expected (not displayed here).

The behavior described above can also be understood considering the pole structure
in the complex $\sqrt{s}$-plane. 
The narrow peak around $1$ GeV for $\lambda=1.0$ corresponds 
to a pole at $s\approx(1.084^{2}-i\hspace{0.02cm}1.084\cdot0.270)$ GeV$^2$, 
which TR has found on the 
second sheet. This pole is indeed present only if $\lambda$ exceeds the critical value 
$\lambda_{c,1}^{\text{TR}} \approx 0.75$. The pole emerges close to (but not on) 
the real axis for $\lambda_{c,1}^{\text{TR}}$
and descends down into the complex plane on the second sheet for increasing coupling 
strength. One can interpret this appearance and motion of a pole as a feature typical
for the kind of dynamical generation we are interested in.
\begin{figure*}[t]
\hspace*{-1.2cm} \begin{minipage}[hbt]{8cm}
\centering
\includegraphics[scale=0.9]{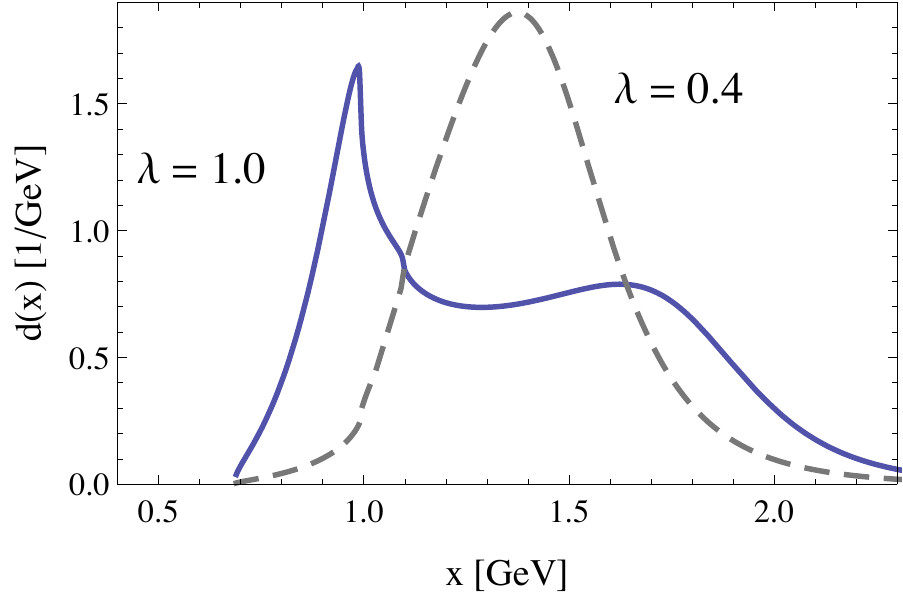}
\end{minipage}
\begin{minipage}[hbt]{8cm}
\centering
\vspace{0.2cm}
\includegraphics[scale=0.969]{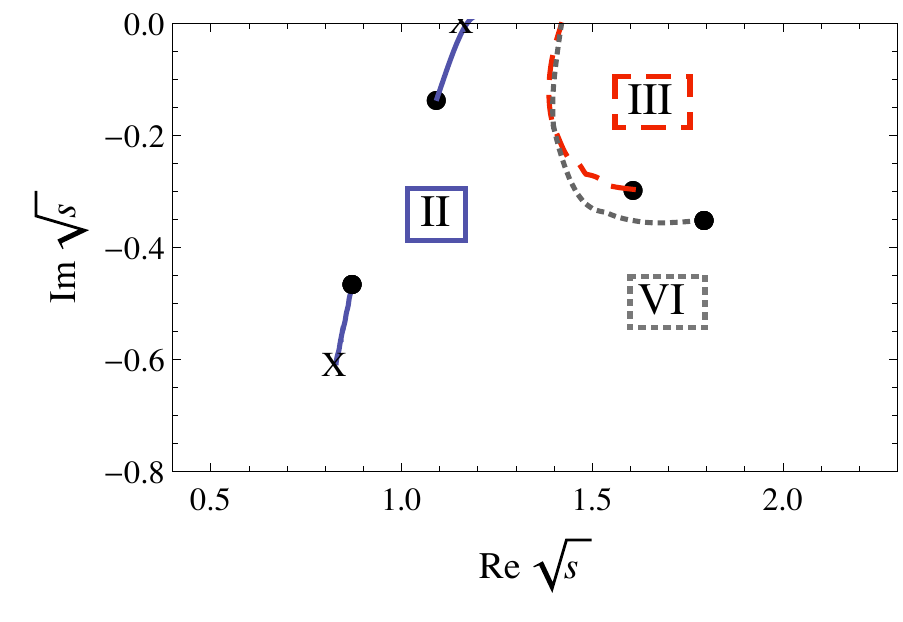}
\end{minipage}
\hspace*{-1.2cm} \begin{minipage}[hbt]{8cm}
\centering
\includegraphics[scale=0.9]{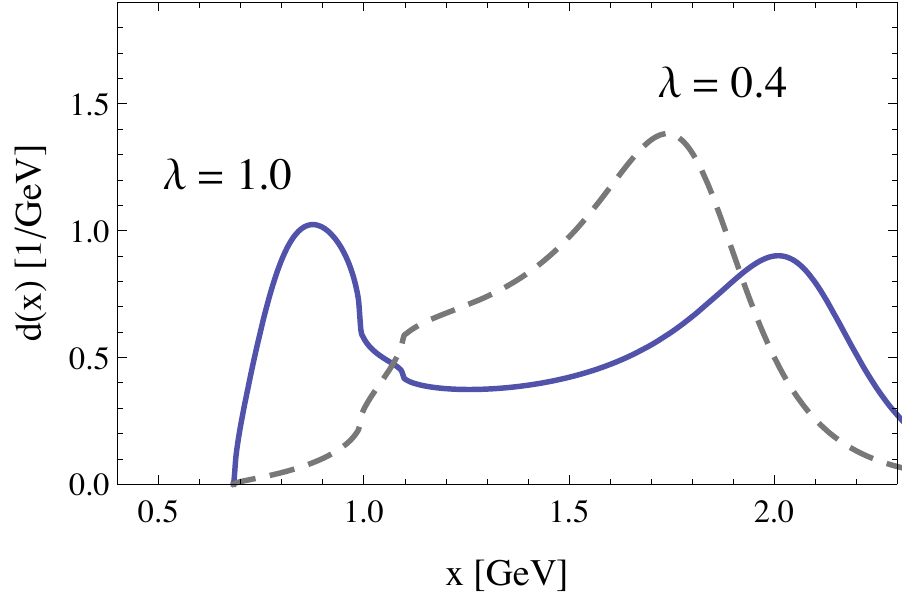}
\end{minipage}
\begin{minipage}[hbt]{8cm}
\centering
\vspace{0.2cm}
\includegraphics[scale=0.969]{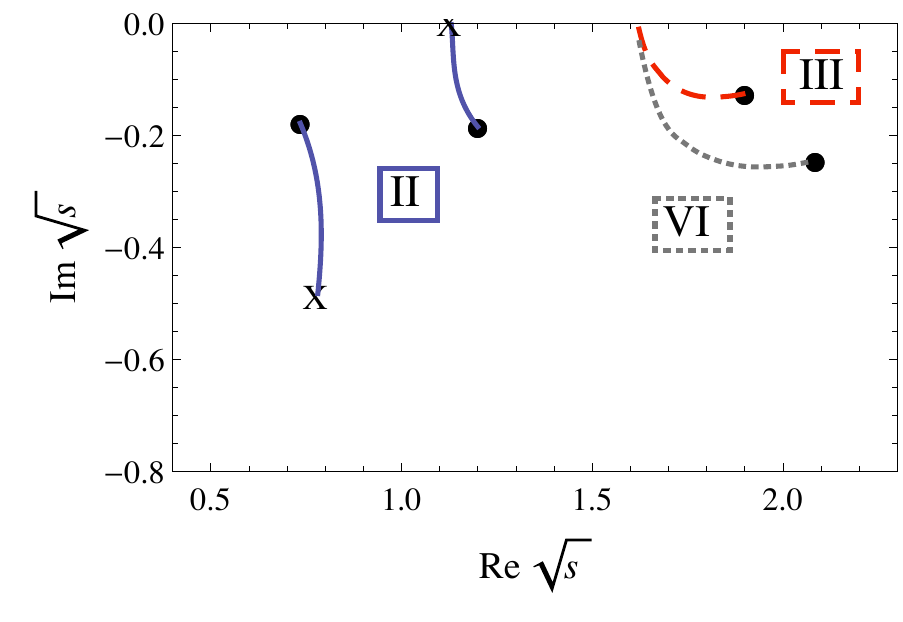}
\end{minipage}
\caption{Spectral functions (left panels) and positions of poles in the complex 
$\sqrt{s}$-plane (right panels) for the parameter sets of
TR (upper row) and BP (lower row). Spectral functions are shown for $\lambda =0.4$ (dashed
grey lines) and $\lambda=1.0$ (solid blue lines). The pole trajectories of
the seed state are indicated by grey dotted or red dashed lines (for details, see the text) and 
the one for the dynamically generated
resonance by solid blue lines. The roman numerals
indicate the Riemann sheets where the respective poles can be found.
Final pole positions $(\lambda =1.0)$ are indicated by solid black dots, pole positions at
$\lambda_{c,i}$, \emph{i.e.}, where the pole $i$ first emerges, are indicated by X.}
\label{fig:spec_and_poles}
\end{figure*}

However, we also find another pole on the second sheet emerging at a
large imaginary value of $\sqrt{s}$ and moving up towards the real axis. 
It first appears for $\lambda_{c,2}^{\text{TR}} \approx0.84$. Its effect on the spectral
function is hard to discern, since (the absolute value of) its imaginary part (\emph{i.e.}, its 
decay width) is still too large at $\lambda=1.0$. This pole was not reported in TR, 
yet, in Ref.\ \cite{comm_beveren}, a similar situation was described where 
the $a_{0}(980)$ was taken to have such a behavior, \emph{i.e.}, its pole was coming 
from the region of large negative imaginary parts of $\sqrt{s}$ and heading towards the 
real axis. This is, however, not the case for the pole of TR, which is dynamically
generated near the real axis and then shifts towards larger (negative) imaginary
values of $\sqrt{s}$.

On the third sheet, TR reports another pole. One could think that this
pole corresponds to the seed state, since for $\lambda=0$ the pole trajectory starts 
on the real axis at the mass of the seed state. However, the pole lies on the third sheet,
\emph{i.e.}, prior to crossing the $\pi\eta^{\prime}$-threshold, but its location is above
that threshold. Therefore, it should not be considered to induce the broad bump 
in the spectral function. However, there is also a pole on the sixth sheet which also 
starts at the mass of the seed state. From its position this pole can also be considered 
to generate the broad resonance shape in the spectrum above $1.5$ GeV. 
It is interesting that the pole on the 
third sheet was suggested in TR to correspond to the $a_{0}(1450)$ resonance. 
From our point of view, because of the above arguments it is more natural to 
take the pole on the sixth sheet. A close inspection of the peak position of the
broad bump in the spectral function reveals that it
corresponds more closely to the real part of the pole on the sixth sheet than that on
the third sheet, which corroborates our interpretation.

\item In the lower row of Fig.\ \ref{fig:spec_and_poles} we present the results for the 
parameter choice in BP. We find that the qualitative behavior is very similar to the one in TR. 
Quantitatively, we find that the bump in the spectral function corresponding to the 
$a_{0}(980)$ resonance is now somewhat wider. The broad structure at large
$\sqrt{s}$ is more pronounced and now lies around $2$ GeV. For decreasing coupling 
strength the peak becomes narrower and moves towards $1.6$ GeV (because the 
seed state is located there).

We find again two poles on the second sheet. 
The right pole appears first for $\lambda_{c,1}^{\text{BP}}\approx0.69$ and the
left one for $\lambda_{c,2}^{\text{BP}}\approx0.66$. The parameter set of BP does 
not yield a pole structure from which one can infer which pole corresponds to the 
$a_{0}(980)$. Both poles give too large widths, and the left one is too light, while the right 
one is too heavy. It seems that both of them are relevant in the generation of the 
bump at $1$ GeV in the spectrum. Moreover, it does not seem to be appropriate
to assign the poles on the other two sheets to $a_{0}(1450)$. At least within this model 
and with the chosen parameters, the pole masses are definitely too high 
\footnote{From the discussion of the poles, we can furthermore
conclude that crossings of the running mass are not really indicative of poles
in the propagator. BP reports three crossings, the first at a mass value close to $1$ GeV,
the second one around $1.4$ GeV, while a third one is located around $1.8$ GeV.
The latter was discarded in BP as unphysical 
[see Ref.\ \cite{boglione} for more details]. From our point of view it is not possible
to unambiguously assign poles to these crossings.}. 
\end{enumerate}

\section{Simple effective model with derivative interactions}
\label{sec:section3}
In the previous section we have re-examined the approach of
TR and BP to dynamically generate resonances in the scalar--isovector sector. 
We now apply the above mechanism of dynamical generation of resonances 
using a formulation based on an interaction Lagrangian.

\subsection{Interactions with derivatives: a lesson from the eLSM}
\label{sec:sec3.1}
The way a scalar field couples to pseudoscalar states depends on the effective 
approach used. Let us, for instance, consider the coupling of $a_{0}$ to kaons.
In chiral perturbation theory (chPT) \cite{chpt,*[see also ][ and refs. therein]chpt2}, 
which is based on the nonlinear realization of chiral symmetry, 
only derivative couplings of the type $a_{0}^{0}\partial_{\mu}K^{0}\partial^{\mu}\bar{K}^{0}$
can appear in the chiral limit \cite{Ecker}. Away
from the chiral limit, a non-derivative coupling $a_{0}^{0}K^{0}\bar{K}^{0}$ appears, too, 
but its strength is proportional to $m_{K}^{2}$, \emph{i.e.}, via the Gell-Mann--Oakes--Renner
relation proportional to the explicit breaking of chiral
symmetry by nonzero quark masses. On the other hand, if the standard
linear sigma model (without vector degrees of freedom) is considered, the coupling is 
only of the non-derivative type $a_{0}^{0}K^{0}\bar{K}^{0}$. 
At tree-level both chPT and the sigma model can coincide, but when loops are included 
differences arise due to the different $s$-dependence in the amplitudes.

Studying the spectral function of 
$\phi\rightarrow a_{0}(980)\gamma\rightarrow\pi^{0}\eta\gamma$ measured by the 
KLOE Collaboration \cite{kloe}, it was shown in Ref.\ \cite{pagliaraderivatives,*pagliaraderivatives2} 
that a derivative coupling of the type 
$a_{0}^{0}\partial_{\mu}K^{0}\partial^{\mu}\bar{K}^{0}$ seems to be necessary. 
As we shall demonstrate below, we come to the same conclusion: 
a derivative coupling is necessary for the simultaneous description of both 
resonances $a_{0}(980)$ and $a_{0}(1450)$.

Interestingly, an improved version of the linear sigma model, called extended 
Linear Sigma Model (eLSM), naturally contains both non-derivative and derivative 
coupling terms. This feature is due to the inclusion of (axial-)vector degrees
of freedom in the model, for more details see Refs.\ \cite{eLSM1,*eLSM1-2,*eLSM1-3,eLSM2}. This 
model is able to provide a surprisingly good description of the tree-level masses and 
decay widths of meson resonances below $1.7$ GeV \cite{eLSM1,*eLSM1-2,*eLSM1-3,eLSM2,stani}. 
Furthermore, in this approach the resonance $a_{0}(1450)$ turns out to be predominantly 
a quark-antiquark state with a (bare) mass of $m_{a_{0}}=1.363$ GeV.

The Lagrangian for the scalar--isovector sector emerging from the eLSM has the 
following form:
\begin{widetext}
\begin{eqnarray}
\mathcal{L}_{a_{0}\eta\pi}^{\text{eLSM}} & = & A_{1}^{\text{eLSM}}a_{0}^{0}\eta\pi^{0}
+B_{1}^{\text{eLSM}}a_{0}^{0}\partial_{\mu}\eta\partial^{\mu}\pi^{0}
+C_{1}^{\text{eLSM}}\partial_{\mu}a_{0}^{0}(\pi^{0}\partial^{\mu}\eta
+\eta\partial^{\mu}\pi^{0}) \ , \label{eq:Lagrangian} \\
\mathcal{L}_{a_{0}\eta'\pi}^{\text{eLSM}} & = & A_{2}^{\text{eLSM}}a_{0}^{0}\eta'\pi^{0}
+B_{2}^{\text{eLSM}}a_{0}^{0}\partial_{\mu}\eta'\partial^{\mu}\pi^{0}
+C_{2}^{\text{eLSM}}\partial_{\mu}a_{0}^{0}(\pi^{0}\partial^{\mu}\eta'
+\eta'\partial^{\mu}\pi^{0}) \ , \nonumber \\
\mathcal{L}_{a_{0}K\bar{K}}^{\text{eLSM}} & = & A_{3}^{\text{eLSM}}
a_{0}^{0}(K^{0}\bar{K}^{0}-K^{-}K^{+})+B_{3}^{\text{eLSM}}a_{0}^{0}
(\partial_{\mu}K^{0}\partial^{\mu}\bar{K}^{0}-\partial_{\mu}K^{-}\partial^{\mu}K^{+}) 
\nonumber \\
&  & + \ C_{3}^{\text{eLSM}}\partial_{\mu}a_{0}^{0}(K^{0}\partial^{\mu}\bar{K}^{0}
+\bar{K}^{0}\partial^{\mu}K^{0}-K^{-}\partial^{\mu}K^{+}-K^{+}\partial^{\mu}K^{-}) \nonumber \ ,
\end{eqnarray}
\end{widetext}
where $A_{i}^{\text{eLSM}}$, $B_{i}^{\text{eLSM}}$, and $C_{i}^{\text{eLSM}}$ are coupling 
constants that are functions of the parameters of the model \cite{eLSM2}. 
Note that both non-derivative 
and derivative interactions appear. The derivatives in front of the fields produce 
an $s$-dependence in the decay amplitudes, $-i\mathcal{M}_{i}^{\text{eLSM}}(s)$, 
which enter the tree-level expressions of the decay widths,
\begin{equation}
\Gamma_{i}^{\text{eLSM}}(s) = \frac{k_{i}(s)}{8\pi s}|\text{--}
i\mathcal{M}_{i}^{\text{eLSM}}(s)|^{2}\Theta(s-s_{th,i}) \ ,
\end{equation}
which have to be evaluated for $s = m_{a_0}^2$. The amplitudes read
\begin{equation}
\mathcal{M}_{i}^{\text{eLSM}}(s) = A_{i}^{\text{eLSM}}
-\frac{1}{2} B_{i}^{\text{eLSM}}\left(s-m_{i1}^{2}-m_{i2}^{2} \right)+C_{i}^{\text{eLSM}}s \ ,
\label{eq:eLSMamp}
\end{equation}
where the masses $m_{i1},m_{i2}$ are the pseudoscalar masses in the relevant channels.
The parameters of the eLSM were determined from a $\chi^2$-fit to tree-level 
masses and decay widths. So far, no loop corrections were considered.
For a consistent loop calculation one would have to perform a new fit of the parameters, 
which is an 
interesting project for future work. In any case, one should 
not use the values of the parameters of the eLSM determined in
Ref.\ \cite{eLSM2} in the expressions for $A_i^{\text{eLSM}},B_i^{\text{eLSM}}$, 
and $C_i^{\text{eLSM}}$. Therefore, we shall treat the latter as free parameters
in the following.

A first attempt to incorporate loop corrections in a scheme inspired by the eLSM was
presented in Refs.\ \cite{procEEF70,proceqcd}. There, the $s$-dependence of the 
amplitudes was completely neglected and a regularization function was introduced,
\begin{equation}
-i\mathcal{M}_{i}^{\text{eLSM}}(s) \ \rightarrow \ -i\mathcal{M}_{i}(s) 
= -i\mathcal{M}_{i}^{\text{eLSM}}(m_{a_{0}}^{2})F_{i}(s) \ .
\end{equation}
After that, the imaginary part of the self-energy was computed using the optical theorem,
\begin{equation}
\operatorname{Im}\Pi_{i}(s) = - \sqrt{s}\, \Gamma_{i}^{\text{tree}}(s) = - \frac{k_{i}(s)}{8\pi 
\sqrt{s}}|\text{--}i\mathcal{M}_{i}(s)|^{2}\Theta(s-s_{th,i}) \ ,
\label{eq:optical}
\end{equation}
and the real part from the dispersion relation (\ref{eq:disp}). As shown in
Ref.\ \cite{procEEF70} 
the model yields a width of the seed state which 
is too small. Moreover, \emph{no} additional pole for the $a_{0}(980)$ is
dynamically generated. Obviously, neglecting the $s$-dependence of the amplitudes 
is an oversimplification. One has to take into account the derivatives in some way; 
at the same time care is needed when derivative interactions appear in a Lagrangian, 
for details see App.\ \ref{app:appendix1}.

\subsection{Effective model with both non-derivative and derivative interactions}

We now consider an effective model for the isovector states containing the same 
decay channels as the eLSM and including also non-derivative and derivative interactions. 
The Lagrangian is given by the sum of the following terms:
\begin{eqnarray}
\mathcal{L}_{a_{0}\eta\pi} & = & A_{1}a_{0}^{0}\eta\pi^{0}
+B_{1}a_{0}^{0}\partial_{\mu}\eta\partial^{\mu}\pi^{0} \ , \label{eq:Lag_eff} \\
\mathcal{L}_{a_{0}\eta^{\prime}\pi} & = & A_{2}a_{0}^{0}\eta^{\prime}\pi^{0}
+B_{2}a_{0}^{0}\partial_{\mu}\eta^{\prime}\partial^{\mu}\pi^{0} \ , \nonumber \\
\mathcal{L}_{a_{0}K\bar{K}} & = & A_{3}a_{0}^{0}(K^{0}\bar{K}^{0}-K^{-}K^{+})
+B_{3}a_{0}^{0}(\partial_{\mu}K^{0}\partial^{\mu}\bar{K}^{0} \nonumber \\
& & - \ \partial_{\mu}K^{-}\partial^{\mu}K^{+}) \nonumber \ .
\end{eqnarray}
Formally it can be obtained by rewriting the terms proportional to $C_{i}^{\text{eLSM}}$ 
in Eq.\ (\ref{eq:Lagrangian}) by an integration by parts in order to get rid of
the derivatives of the $a_0$-fields \footnote{Quite remarkably, terms with derivatives acting 
on the decaying field are pretty peculiar, see App.\ \ref{app:appendix1} for some remarks.}. 
Subsequently, one replaces the emerging second 
derivatives with the help of the Klein--Gordon equation, $\square\pi^{0}=-m_{\pi}^{2}\pi^{0}$
(and similarly for the other pseudoscalar fields). Then, Eq.\ (\ref{eq:Lag_eff})
gives rise to the following $s$-dependent amplitudes:
\begin{equation}
\mathcal{M}_{i}^{\text{eff}}(s) = \left[A_{i}-\frac{1}{2} B_{i}\left(s-m_{i1}^{2}-m_{i2}^{2}\right)
\right] F_{i}(s) \ ,
\end{equation}
where we have already included a regularization function $F_{i}(s)$ as defined in Sec.\
\ref{sec:section2}. We again note that the constants $A_{i}$ and $B_{i}$ will not
be computed from the numerically determined parameters of the eLSM, but
will be determined in order to produce the masses and decay widths of the resonances
under study. Note that in chPT the parameters
$A_{i}$ are proportional to the masses of the pseudo-Goldstone bosons as
$A_{1}\propto m_{\pi}^{2}+m_{\eta}^{2}$, $A_{2}\propto m_{\pi}^{2}
+m_{\eta^{\prime}}^{2}$, $A_{3}\propto2m_{K}^{2}$ and thus vanish in the chiral limit. Thus, also from this
consideration, we expect that the derivative terms are sizable and crucial for
the determination of the resonance poles.

We computed the real and imaginary part of the self-energy in two ways.
In the first, we applied the method outlined in Sec.\ \ref{sec:sec3.1}, \emph{i.e.}, 
we computed the
tree-level decay widths and used the optical theorem from Eq.\ (\ref{eq:optical}) 
to obtain the imaginary
part of the self-energy. We then applied the dispersion relation (\ref{eq:disp})
to calculate the
corresponding real part. In the second approach, we computed the one-loop
self-energy directly from the Feynman rules. From a comparison, we identified 
the necessity to introduce subtractions in the first approach, for details see
App.\ \ref{app:appendix1}. Note that the one-loop approximation for the self-energy is
quite reliable, since vertex corrections can be shown to have a negligible effect \cite{jonas}.

There are eight parameters in our approach: $m_{0},\ \Lambda=\sqrt{2}k_{0},$ and
six coupling constants $A_i,B_i$ $(i=1,2,3)$. We vary the numerical values
of $m_{0}$ and $\Lambda$ within reasonable intervals $m_{0} \in (0.8,1.5)$ GeV
and $\Lambda \in (0.4,1.5)$ GeV and each time perform a fit of the six
coupling constants to six experimental quantities: one pole in the PDG range
for $a_{0}(980) \ (\text{in our case }\sqrt{s}=(0.969-i\hspace{0.02cm}0.045) \ \text{GeV})$ and one for 
$a_{0}(1450) \ (\text{in our case }\sqrt{s}=(1.450-i\hspace{0.02cm}0.135) \ \text{GeV})$, and 
the central values of the branching ratios of $a_{0}(1450)$ [see Eq.\ (\ref{eq:branching})]. 
By this, all six free parameters can be fixed.

\begin{figure*}[t]
\hspace*{-1.5cm}
\begin{minipage}[hbt]{8cm}
\centering
\includegraphics[scale=0.9]{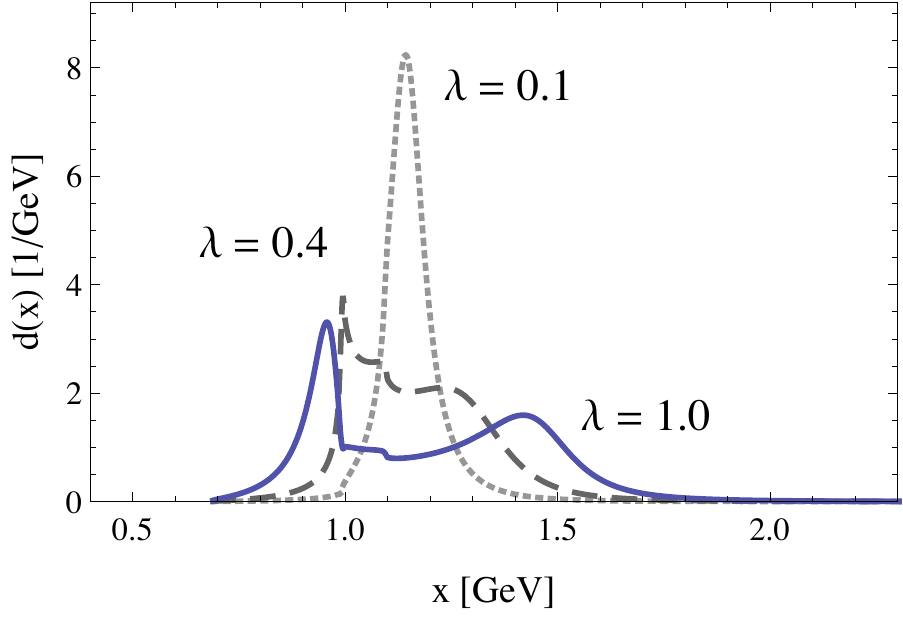}
\end{minipage}
\hspace*{0.2cm}
\begin{minipage}[hbt]{8cm}
\centering
\vspace*{0.2cm}
\includegraphics[scale=1.02]{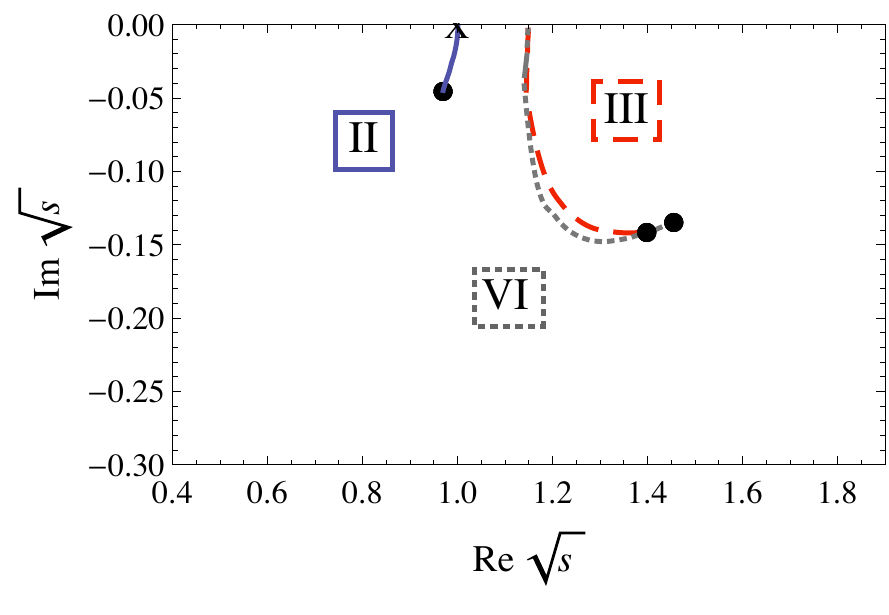}
\end{minipage}
\caption{In the left panel we
show the spectral functions for three different values of $\lambda$. In the right panel
we display pole trajectories obtained by varying $\lambda$ from zero to one. 
Black dots indicate the position of the poles for $\lambda=1.0$. The X indicates the
pole position for $\lambda_c$, \emph{i.e.}, when the pole first emerges.
The roman numeral indicates on which sheet the respective pole can be found.}
\label{fig:toy}
\end{figure*}

It turns out that there is only a \textit{narrow} range of suitable values of
the parameters $m_{0}$ and $\Lambda$ for which the fit of the six coupling
constants is possible: approximately $m_{0} \in (0.9,1.2)$ GeV and $\Lambda \in (0.4,0.9)$ GeV. Here, 
`approximately' refers to 
the fact that, due to the interdependence of the parameters, the window is not rectangular. 
However, a small change in $m_{0}$ and/or $\Lambda$ by 50 MeV near the borders of the quoted interval
does not allow one to reproduce the data anymore. Thus, although we have eight parameters, 
we are severely constrained in their choice in order to describe the $I=1$ resonance. 
As we will see below, the present parameters also explain why $a_{0}(980)$ couples strongly to kaons.
The final values for
the parameters and coupling constants are:
\begin{align}
m_{0}&=1.15 \ \text{GeV} \ , & \Lambda &= 0.6 \ \text{GeV} \ , \\
A_{1} &= 2.52 \ \text{GeV} \ , & B_{1} &= -8.07 \ \text{GeV}^{-1} \ , \label{eq:AiBi} \\
A_{2} &= 9.27 \ \text{GeV} \ , & B_{2} &= 9.25 \ \text{GeV}^{-1} \ , 
\nonumber \\
A_{3} &= -6.56 \ \text{GeV} \ , & B_{3} &= -1.54 \ \text{GeV}^{-1} \ . \nonumber
\end{align}
We rescale these coupling constants by a common factor $\sqrt{\lambda}$ 
and compute the corresponding spectral functions. 
The result is shown in the left panel of Fig.\ \ref{fig:toy}. We also compute the
pole trajectories in the complex $\sqrt{s}$-plane by varying $\lambda$ from zero to one.
The following comments are in order:
\begin{enumerate}
\item The spectral function shows a narrow peak for $\lambda=1.0$ at a value of 
$x= \sqrt{s}$ slightly smaller than $1$ GeV, which can be interpreted as the $a_{0}(980)$. 
The form is distorted by the nearby $K\bar{K}$-threshold and resembles the Flatt\'{e} 
distribution \cite{flatte,*flatte2}, see also Refs.\ \cite{giacosapagliara,baru} and refs.\ therein. 
The pole corresponding to this peak lies on the second sheet and has coordinates
\begin{equation}
\sqrt{s} = (0.970-i\hspace{0.02cm}0.045) \ \text{GeV} \ ,
\end{equation}
\emph{i.e.}, we find the $a_{0}(980)$ to have a mass of $m_{\text{pole}}=0.969$ GeV 
and a width of $\Gamma_{\text{pole}}=0.090$ GeV. 
This pole appears only if $\lambda$ exceeds $\lambda_c\approx0.52$ (note that the pole trajectory is very different 
from the one reported in Ref. \cite{zhiyong}). The corresponding
position is indicated by an X in the right panel of Fig.\ \ref{fig:toy}. The important thing here 
is that, in contrast to what we have found for the TR and BP parametrizations, there is 
only one pole for the $a_{0}(980)$, and thus no ambiguity which one
should be identified with this resonance \footnote{There are two additional poles in the 
relevant part of the complex plane which will not be displayed and discussed here: 
$(i)$ a pole deep in the imaginary region on the second sheet, and $(ii)$ a pole close
to the imaginary axis on the sixth sheet. Both have no physical impact. 
Quite interestingly, we do not observe any virtual bound states. 
Such poles were described by two of us within 
theories without derivative interactions, \emph{e.g.} in Refs.\ \cite{e38,thomasthesis}.}.

\item There is also a broad structure around $1.5$ GeV which corresponds to the resonance 
$a_{0}(1450)$. For decreasing $\lambda$, both peaks merge and settle around 
$1.15$ GeV, where the seed state is located.

\item As expected, there is (only) one pole present on the third sheet with 
coordinates $\sqrt{s}=(1.400-i\hspace{0.02cm}0.141)$ GeV. However, as in 
TR and BP, we find a pole on the sixth sheet, too. Its coordinates are
\begin{equation}
\sqrt{s} = (1.456-i\hspace{0.02cm}0.134)  \ \text{GeV} \ ,
\end{equation}
or $m_{\text{pole}}=1.450$ GeV and $\Gamma_{\text{pole}}=0.270$ GeV. 
This is the pole which is responsible for the peak around $1.5$ GeV in the spectrum, and thus
we assign it to the $a_{0}(1450)$. However, since the pole on the third sheet also
reproduces mass and width of $a_{0}(1450)$ to reasonable accuracy, 
it is in principle possible to regard this one 
as the pole corresponding to $a_{0}(1450)$, too.
\end{enumerate}

The present study demonstrates that, by starting with a unique seed state, it is indeed 
possible to find two poles for the isovector states, both of which reproduce the
masses and widths of $a_0(980)$ and $a_0(1450)$ reasonably well.

\subsection{Branching ratios and coupling constants for $a_{0}(980)$}
For completeness, we report the branching ratios of our effective model by using 
the tree-level decay widths obtained from the optical theorem (\ref{eq:optical}). 
The partial widths are evaluated at the peak value of the spectral function above $1$ GeV, 
$m_{\text{peak}}=1.419$ GeV. For the resonance $a_{0}(1450)$ this leads to
\begin{equation}
\frac{\Gamma_{a_{0}\rightarrow\eta^{\prime}\pi}^{\text{tree}}}{
\Gamma_{a_{0}\rightarrow\eta\pi}^{\text{tree}}}\simeq0.44 \ , \ \ \ 
\frac{\Gamma_{a_{0}\rightarrow K\bar{K}}^{\text{tree}}}{
\Gamma_{a_{0}\rightarrow\eta\pi}^{\text{tree}}}\simeq0.96 \ ,
\end{equation}
which can be compared to the experimental values \cite{olive}:
\begin{equation}
\frac{\Gamma_{a_{0}\rightarrow\eta^{\prime}\pi}}{\Gamma_{a_{0}\rightarrow\eta\pi}}
=0.35\pm0.16 \ , \ \ \ \frac{\Gamma_{a_{0}\rightarrow K\bar{K}}}{\Gamma_{a_{0}\rightarrow\eta\pi}}
=0.88\pm0.23 \ .
\label{eq:branching}
\end{equation}

Concerning the resonance $a_{0}(980)$, we give the following estimates for the coupling 
constants in the $\pi\eta$- and $K\bar{K}$-channels: We calculate the partial widths 
$\Gamma_i^{\text{tree}}(s)$, this time with $\sqrt{s}$ equal to 
the peak mass of the spectral function 
below $1$ GeV, $m_{\text{peak}}=0.956$ GeV. Then, Eq.\
(\ref{eq:optical}) is used to solve for the absolute values of the amplitudes. The result
is multiplied with the root of the wave-function renormalization factor, $\sqrt{Z}=0.652$, 
which is its value at the Breit--Wigner mass of the $a_{0}(980)$. Thus we obtain
the coupling constants in the $\pi\eta$- and $K\bar{K}$-channels as
\begin{equation}
g_{\pi\eta}=2.496 \ \text{GeV} \ , \ \ \ g_{K\bar{K}}=6.012 \ \text{GeV} \ .
\end{equation}
It is remarkable that the coupling of $a_{0}(980)$ to kaons turns out to be sizably larger 
than the coupling to $\pi\eta$. This is in agreement with various other works on this topic
\cite{molecular,*molecular2,*molecular3,*molecular4,*molecular5,pagliaraderivatives,*pagliaraderivatives2,maiani}: 
Virtual kaon-kaon pairs near the kaon-kaon 
threshold are important for the dynamical generation of the resonance $a_{0}(980)$.

\section{Conclusions}
\label{sec:conc}
\begin{table*}[t]
\begin{ruledtabular}
\begin{tabular}{lcccc}
\multicolumn{3}{c}{\hspace{3.5cm}$a_{0}(980)$} & \multicolumn{2}{c}{$a_{0}(1450)$}\\[0.1cm]
& $m_{\text{pole}}$ [GeV] & $\Gamma_{\text{pole}}$ [GeV] & $m_{\text{pole}}$ [GeV] 
& $\Gamma_{\text{pole}}$ [GeV]\\[0.05cm]
\hline\\
TR \cite{tornqvist,*tornqvist2} & $1.084$ & $0.270$ & $1.566$ & $0.578$\\
BP \cite{pennington} & $1.186$* & $0.373$* & $1.896$ & $0.250$\\
Our results \ & $0.969$ & $0.090$ & $1.450$ & $0.270$\\
&  &  &  &\\
\vspace{0.3cm}PDG \cite{olive} & $ \ 0.980\pm0.020 \ $ & $ \ 0.050$ to $0.100$ \ & $ \ 
1.474\pm0.019 \ $ 
& $ \ 0.265\pm0.013$\\
\multicolumn{5}{l}{\footnotesize{*In order to compare to TR, the right pole on the second 
sheet was chosen.}}\\
\end{tabular}
\end{ruledtabular}
\caption{Numerical results for the pole coordinates in the scalar--isovector sector in 
TR, BP, and our effective model, compared to the PDG values. In the case of the 
$a_{0}(1450)$, the poles listed for TR and BP are located on the third sheet, while our pole 
lies on the sixth sheet. All poles for the $a_{0}(980)$ are found on the second sheet. 
Note that all poles listed for BP were obtained performing the analytic 
continuation of the propagator given by BP.}
\label{tab:tab2}
\end{table*}

Experimental data exhibit several puzzling facts about the light scalar mesons: 
$f_{0}(500)$ (or $\sigma$) and $K_{0}^{\ast}(800)$ have 
large decay widths, while $f_{0}(980)$ and $a_{0}(980)$ are narrow 
but their spectral functions show threshold distortions due to the nearby $K\bar{K}$-threshold. 
It is nowadays recognized that these states do not fit into the ordinary $q\bar{q}$ picture 
based on a simple representation of $SU(3)$--flavour symmetry \cite{isgur}, yet there is 
no consensus on the precise mechanism which generates them. One can also regard 
these states as four-quark objects, for example as tetraquarks 
\cite{jaffe,*jaffe2,maiani,fariborz,*fariborz2,rodriguez,tqmix} or as dynamically generated states. 
The latter are states which are not present in the original formulation of a hadronic model 
but appear when calculating loop corrections
\cite{oller,*oller2,*oller3,pelaez,*pelaez2,dullemond,tornqvist,*tornqvist2,morgan,boglione}. Indeed, the interpretation of 
light scalar states as loosely bound molecular states 
\cite{molecular,*molecular2,*molecular3,*molecular4,*molecular5} is also an example 
of dynamical generation. [For another interpretation of light scalar states, see \emph{e.g.} 
Ref.\ \cite{ochs,*ochs2,*ochs3}.]

A particular type of dynamical generation is that of `image' or `companion poles'. 
We have concentrated on such a method in this work and have applied it to the 
scalar--isovector sector. Our results demonstrate that it is in fact possible to correctly 
describe the resonances $a_{0}(980)$ and $a_{0}(1450)$ in a unique framework, 
where originally only a single quark-antiquark seed state is present.

Besides that, we have also repeated the previous calculations of 
T\"ornqvist and Roos \cite{tornqvist,*tornqvist2} and Boglione and Pennigton \cite{pennington}. 
These studies have been extended by us to the complex plane on all Riemann sheets 
nearest to the first, physical sheet. A summary of our results and, for comparison, 
those of T\"ornqvist and Roos, and Boglione and Pennington, can be found in Tab.\
\ref{tab:tab2}. Our results are based on an effective Lagrangian approach that includes 
both derivative and non-derivative interaction terms, see Eq.\ (\ref{eq:Lag_eff}), inspired by 
the extended Linear Sigma Model (eLSM), and show that both terms are necessary 
and equally important \cite{eLSM1,*eLSM1-2,*eLSM1-3,eLSM2}.

Note that the formulation of dynamical generation applied here is related, but not 
equal to the one described in Ref.\ \cite{oller,*oller2,*oller3}. In the latter the scattering amplitude
is computed from an effective Lagrangian (derived from chiral perturbation theory
and containing only pseudoscalar mesons) and then unitarized; this process of 
unitarization generates, for instance, the pole of the $a_{0}(980)$ in the 
scalar--isovector sector. Yet, it is $a\ priori$ not possible to know if the resulting state is 
in fact a quark-antiquark or a four-quark resonance, and if it can be linked to the 
heavier $a_{0}(1450)$ state or not; see Ref.\ \cite{giacosaDynamical} for a detailed 
discussion of this issue. However, further studies within this scheme were performed in Ref.\
\cite{oller4} by including an octet of bare resonances masses
around $1.4$ GeV. It was found that the physical
$a_{0}(1450)$ in fact originates from this octet, giving a clear statement
about its nature, which is in agreement with our results.

In this work we have concentrated on the isovector sector $I=1.$ However, the
very same mechanism is applicable for the other light and heavy scalar states.
As a consequence, all light scalars are dynamically generated states. In
particular, it seems promising to extend the present study in the low-energy
regime into two directions: $(i)$ The isodoublet, \emph{i.e.}, by describing the resonances 
$K_{0}^{\ast}(800)$ and $K_{0}^{\ast}(1430)$ in a similar framework. 
The pole of $K_{0}^{\ast}(800)$ is not yet very well known and there is need of 
improved analyses from different directions. $(ii)$ The scalar--isoscalar sector, 
where the resonances $f_{0}(500)$ and $f_{0}(980)$ should be dynamically generated.
In this case, $f_{0}(1370),$ $f_{0}(1500)$, and $f_{0}(1710)$ would be predominantly 
a non-strange quarkonium, a strange quarkonium, and a scalar glueball, respectively.

Another interesting subject is the study of dynamical generation in the framework of 
puzzling resonances in the charmonium sector \cite{PenningtonCharm}, see for example Ref.\ \cite{brambilla} 
and refs.\ therein. Namely, a whole class of mesons, called $X$, $Y$, and $Z$ states, 
has been experimentally discovered but is so far not fully understood \cite{braaten,*braaten2,maianix}.
As demonstrated in Ref.\ \cite{coitox} for the case of $X(3872)$, some of the $X$ and $Y$ states 
could emerge as companion poles of quark-antiquark states.

\begin{acknowledgements}
The authors thank M.\ Pennington, J.\ Wambach, G.\ Pagliara, J.\ Reinhardt, D.D.\ Dietrich, 
R.\ Kami\'{n}ski, J.R.\ Pel\'{a}ez, and H.\ van Hees for useful discussions. T.W.\ 
acknowledges financial support from HGS-HIRe, F\&E GSI/GU, and HIC for FAIR Frankfurt.
\end{acknowledgements}

\appendix
\section{}
\label{app:appendix1}

In this Appendix, we compute the one-loop self-energy in the case of derivative interactions.
We first derive the interacting part of the Hamiltonian from the Lagrangian via
a Legendre transformation. We shall see that the derivative interactions give
rise to new interaction vertices. We demonstrate that, in a perturbative calculation of the 
one-loop self-energy, these terms are
necessary to cancel additional terms arising from contractions of
gradients of fields. This proves that, at least at the one-loop level, it is justified to apply
standard Feynman rules with the derivative interaction in ${\cal L}_{\text{int}}$.
We shall also demonstrate that a computation of the
self-energy via the dispersion relation (\ref{eq:disp}) may require subtraction
constants to agree with the perturbative calculation using Feynman rules.

\subsection*{Canonical quantization}

Let us consider a theory with two scalar fields, $S$ and $\phi$, which allows for the 
decay process $S\rightarrow\phi\phi$. Consequently, the Lagrangian is
\begin{equation}
\mathcal{L} = \mathcal{L}_S + \mathcal{L}_\phi + \mathcal{L}_{\text{int}}\;,
\end{equation}
where
\begin{eqnarray}
\mathcal{L}_S & = & \frac{1}{2} \left( \partial_\mu S \partial^\mu S - M^2 S^2 \right)\;,
\nonumber \\
\mathcal{L}_\phi & = & \frac{1}{2} \left( \partial_\mu \phi \partial^\mu \phi - m^2 \phi^2 \right)\;,
\nonumber \\
\mathcal{L}_{\text{int}}& =& gS\partial_{\mu}\phi\partial^{\mu}\phi\;. \label{eq:Lint}
\end{eqnarray}
For perturbative calculations of $\hat{S}$-matrix elements or Green's functions, however, 
one needs the interaction part of the Hamilton operator in the interaction picture. 
We derive this operator via a Legendre transformation of ${\cal L}$ 
and subsequent canonical quantization in the interaction picture.
As a byproduct of this calculation we will explicitly show that the derivative interactions 
invalidate the commonly used relation 
$\mathcal{H}_{\text{int}}=-\mathcal{L}_{\text{int}}$ \cite{reinhardt}. 

The canonically conjugate fields are 
\begin{eqnarray}
\pi_{S} & = & \frac{\partial\mathcal{L}}{\partial(\partial_{0}S)} = \partial^{0}S \ , \\
\pi_{\phi} & = & \frac{\partial\mathcal{L}}{\partial(\partial_{0}\phi)} 
= \partial^{0}\phi+2gS\partial^{0}\phi = (1+2gS) \partial^0 \phi \nonumber \ .
\end{eqnarray}
The Hamiltonian is defined via a Legendre transformation of ${\cal L}$,
\begin{eqnarray}
\mathcal{H} & = & \pi_{S}\partial^{0}S+\pi_{\phi}\partial^{0}\phi-\mathcal{L} \\
& = & \frac{1}{2}\pi_{S}\pi_{S} +\frac{1}{2}\vec{\nabla}S\cdot\vec{\nabla}S
+\frac{1}{2}M^{2}S^{2}
+\frac{1}{2}\pi_{\phi}\pi_{\phi}(1+2gS)^{-1} \nonumber \\
& & + \ \frac{1}{2}\vec{\nabla}\phi\cdot\vec{\nabla}\phi
+\frac{1}{2}m^{2}\phi^{2}+gS\vec{\nabla}\phi\cdot\vec{\nabla}\phi \nonumber \ .
\end{eqnarray}
For a perturbative calculation, we need to expand the denominator 
$(1+2gS)^{-1}$ and obtain the interaction part of the Hamiltonian as
\begin{equation}
\mathcal{H}_{\text{int}}=-gS\pi_{\phi}\pi_{\phi}+gS\vec{\nabla}\phi\cdot\vec{\nabla}\phi
+2g^{2}S^{2}\pi_{\phi}\pi_{\phi}+\mathcal{O}(g^{3}) \ .
\end{equation}
We may now quantize in the Heisenberg picture (indicated by a superscript $H$
at the respective operators). This is commonly done by
promoting fields to operators $S \rightarrow\hat S^{H}$, $\phi\rightarrow\hat\phi^{H}$, 
$\pi_{S}\rightarrow\hat\pi_{S}^{H}$, $\pi_{\phi}\rightarrow\hat\pi_{\phi}^{H}$,
and postulating certain commutation relations for these operators.
However, in perturbation theory we need the operators in the interaction picture.
The following relations hold,
\begin{align}
\hat S^{I} &= \hat U \hat S^{H} \hat U^\dagger \ , & \hat\phi^{I} &= \hat U \hat \phi^{H} \hat U^\dagger \ , \\
\hat \pi_S^{I} &= \hat U \hat \pi_S^{H} \hat U^\dagger \ , & \hat \pi_\phi^{I} &= \hat U \hat \pi_\phi^{H} \hat U^\dagger \nonumber \ ,
\end{align}
where $\hat U=e^{i\hat H_{0}t} e^{-i\hat Ht}$ is the time-evolution
operator that relates operators in the Heisenberg picture with those in the
interaction picture.
Finally, replacing $\hat\pi^{I}_{S} = \partial^{0}\hat S^{I} , \, \hat\pi^{I}_{\phi} 
= \partial^{0}\hat\phi^{I}$ this results in
\begin{equation}
\hat{\mathcal{H}}_{\text{int}}^{I}=-\hat{\mathcal{L}}_{\text{int}}^{I}
+2g^{2}\hat S^{I}\hat S^{I}\partial_{0}\hat\phi^{I}\partial^{0}\hat\phi^{I}+\mathcal{O}(g^{3}) \ .
\label{eq:Hint}
\end{equation}
As advertised, the second term spoils the standard relation 
$\hat{\mathcal{H}}_{\text{int}}=-\hat{\mathcal{L}}_{\text{int}}$. This term
corresponds to a four-point vertex, so it will not appear in the tree-level decay 
$S \rightarrow \phi \phi$. In contrast, in the one-loop self-energy, it will give rise
to an additional tadpole contribution. 

\subsection*{Perturbative calculation of the one-loop self-energy}

We now turn to the self-energy $\Pi(s)$ of the field $S$. At one-loop level,
the Feynman rules applied to $\hat{\cal H}_{\text{int}}$ tell us that 
we will have two contributions. The first contribution
comes from taking two three-point vertices of $\hat{\cal L}_{\text{int}}$ where the $\hat{\phi}$
legs are joined in a manner which gives a 1PI diagram. A covariant derivative acts
on each $\hat{\phi}$ leg at each vertex. The second contribution is a tadpole term
arising from the four-point vertex in Eq.\ (\ref{eq:Hint}), which has two time derivatives 
on the internal leg. This can be graphically depicted as follows:
\begin{equation} \label{eq:Pigraph}
\includegraphics[scale=0.49]{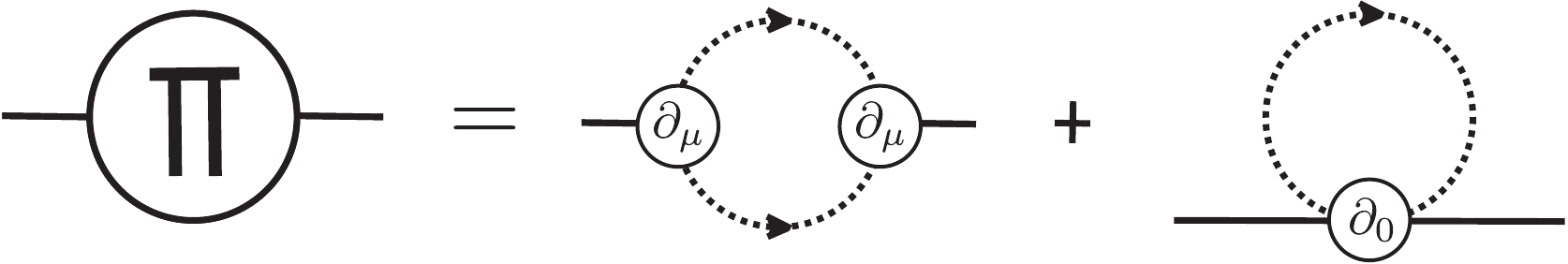}
\end{equation}
The usual Feynman propagator is defined as a contraction of two fields:
\begin{eqnarray} \label{eq:Feynmanprop}
\bcontraction{}{\hat{\phi}}{(x_{1})}{\hat{\phi}}\hat{\phi}(x_{1})\hat{\phi}(x_{2}) & = &
\langle0|\mathcal{T}\big{\{}\hat{\phi}(x_{1})\hat{\phi}(x_{2})\big{\}}|0\rangle \\
& = & \langle0|\hat{\phi}(x_{1})\hat{\phi}(x_{2})|0\rangle\Theta(x_{1}^{0}-x_{2}^{0}) \nonumber \\
& & + \ \langle0|\hat{\phi}(x_{2})\hat{\phi}(x_{1})|0\rangle\Theta(x_{2}^{0}-x_{1}^{0}) \nonumber \\ 
& = & i\Delta_{F}^{\phi}(x_{1}-x_{2}) \nonumber \\
& = & i\int\frac{\text{d}^{4}p}{(2\pi)^{4}}\frac{e^{-ip\cdot(x_{1}-x_{2})}}{p^{2}-m^{2}+i\epsilon} \ . 
\nonumber
\end{eqnarray}
However, in the tadpole diagram, we have the contraction of two fields
on each of which acts a time derivative. Because time-ordering has no
effect at the same space-time point, we obtain:
\begin{eqnarray}  \label{eq:tadpole1}
\langle0|\mathcal{T}\big{\{}\partial_{0}^{x}\hat{\phi}(x)\partial^{0,x}\hat{\phi}(x)\big{\}}|0\rangle
& = & \langle0|{}\partial_{0}^{x}\hat{\phi}(x)\partial^{0,x}\hat{\phi}(x)|0\rangle \nonumber \\
& = & i \int\frac{\text{d}^{4}p}{(2\pi)^{4}}\frac{E_{\textbf{p}}^{2}}{p^{2}-m^{2}+i\epsilon} \ . \nonumber \\
\end{eqnarray}
In order to obtain this result we inserted the standard Fourier decomposition of the
field operators
\begin{equation}
\hat{\phi} (x) = \int\frac{\text{d}^{3}p}{(2\pi)^{3}}\frac{1}{\sqrt{2E_{\textbf{p}}}}
\Big(\hat{a}_{\textbf{p}}^{\dagger}e^{ip\cdot x}+\hat{a}_{\textbf{p}}e^{-ip\cdot x}\Big)\;,
\end{equation}
where $p^0 = E_{\bf p} = \sqrt{{\bf p}^2 +m^2}$ is the on-shell energy.
Thus, a time derivative acting
on a field operator brings down a factor of $\pm i$ times the on-shell energy in the 
corresponding Fourier representation. 

The result (\ref{eq:tadpole1}) is, however, identical if we just act with
the time derivatives on the standard Feynman propagator (\ref{eq:Feynmanprop}):
\begin{eqnarray} \label{eq:tadpole2}
\partial_{0}^{x}\partial^{0,x} \langle0|{}\hat{\phi}(x)\hat{\phi}(x)|0\rangle
& = & \lim_{x_1\rightarrow x_2} \partial_{0}^{x_1}\partial^{0,x_2} 
\langle0|\mathcal{T}\big{\{}\hat{\phi}(x_1)\hat{\phi}(x_2)\big{\}}|0\rangle \nonumber \\
& = & i \int\frac{\text{d}^{4}p}{(2\pi)^{4}}\frac{E_{\textbf{p}}^{2}}{p^{2}-m^{2}+i\epsilon} \ .
\end{eqnarray}
In order to prove this, it is convenient to first perform the $p_0$ integration in
Eq.\ (\ref{eq:Feynmanprop}) and then take the time derivatives.
The equivalence of Eqs.\ (\ref{eq:tadpole1}) and (\ref{eq:tadpole2}) is graphically depicted as
\begin{eqnarray} \label{eq:Pi12}
\langle0|\mathcal{T}\big{\{}\partial_{0}^{x}\phi& &(x)\partial^{0,x}\phi(x)\big{\}}|
0\rangle \ \ \sim \\
& & \includegraphics[scale=0.5]{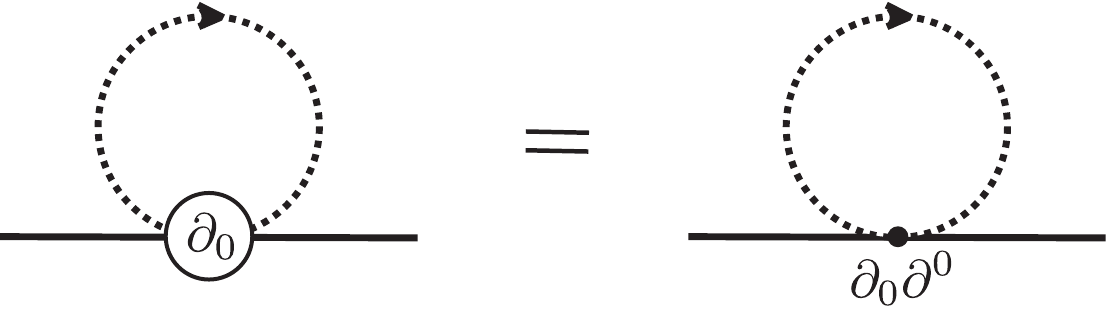} \nonumber
\end{eqnarray}
In the perturbative series of the full propagator of the $S$-field, this tadpole contribution
appears in combination with two free $S$-field propagators (where we omit the
superscript $S$):
\begin{widetext}
\begin{align}
\frac{-2\hspace{0.01cm}i\hspace{0.01cm}g^2}{1!}\cdot2 \int\text{d}x^{\prime} \ i\Delta(x_{1}-x^{\prime})
\langle0|\mathcal{T}\big{\{}\partial_{0}^{x^{\prime}}\phi(x^{\prime})\partial^{0,x^{\prime}}
\phi(x^{\prime})\big{\}}|0\rangle\hspace{0.02cm}i\Delta(x^{\prime}-x_{2}) \\
= \ i\hspace{0.02cm}2g^{2}\cdot2\int\text{d}x^{\prime} \ \Delta(x_{1}-x^{\prime})
\Delta(x^{\prime}-x_{2})\langle0|\mathcal{T}\big{\{}\partial_{0}^{x^{\prime}}\phi(x^{\prime})
\partial^{0,x^{\prime}}\phi(x^{\prime})\big{\}}|0\rangle \nonumber \ .
\label{eq:tadpole_pos}
\end{align}
The factor $-2i g^2$ is the factor accompanying the four-point vertex, 
cf.\ Eq.\ (\ref{eq:Hint}).
A factor of two arises because each $S$ propagator can be joined
with either one of the $S$ legs at the vertex.

We now compute the first diagram in Eq.\ (\ref{eq:Pigraph}). 
To this end, we need contractions of gradients of the $\phi$-fields. These
can be expressed in terms of gradients acting on the standard
Feynman propagator.
The gradient of the Feynman propagator (\ref{eq:Feynmanprop}) is
\begin{eqnarray}
i\hspace{0.02cm}\partial_{\nu}^{x_{2}}\Delta_{F}^{\phi}(x_{1}-x_{2}) & = & \partial_{\nu}^{x_{2}}
\langle0|\mathcal{T}\big{\{}\phi(x_{1})\phi(x_{2})\big{\}}|0\rangle \\
& = & \langle0|\mathcal{T}\big{\{}\phi(x_{1})\partial_{\nu}^{x_{2}}\phi(x_{2})\big{\}}|0\rangle-\eta_{\nu0}\delta(x_{1}^{0}-x_{2}^{0}) 
\langle0|\underbrace{\big[\phi(x_{1}),\phi(x_{2})\big]}_{=0}|0\rangle \ , \nonumber
\end{eqnarray}
where we used the explicit definition of the time-ordered product.
The last term vanishes on account of the delta function, since it is an equal-time commutator
of two $\phi$-fields \cite{reinhardt}.
Taking another gradient leads to
\begin{eqnarray}
i\hspace{0.02cm}\partial_{\mu}^{x_{1}}\partial_{\nu}^{x_{2}}\Delta_{F}^{\phi}(x_{1}-x_{2}) 
& = & \partial_{\mu}^{x_{1}}\langle0|\mathcal{T}\big{\{}\phi(x_{1})\partial_{\nu}^{x_{2}}
\phi(x_{2})\big{\}}|0\rangle \\
& = & \langle0|\mathcal{T}\big{\{}\partial_{\mu}^{x_{1}}\phi(x_{1})\partial_{\nu}^{x_{2}}
\phi(x_{2})\big{\}}|0\rangle+\eta_{\mu0}\delta(x_{1}^{0}-x_{2}^{0})\langle0|\underbrace{\big[\phi(x_{1}),
\partial_{\nu}^{x_{2}}\phi(x_{2})\big]}_{\neq0}|0\rangle \nonumber \ .
\end{eqnarray}
Now the second term does not vanish if $\nu=0$, because then
it involves a commutator of a field with its canonically conjugate field \cite{reinhardt},
\begin{equation}
\eta_{\mu0}\delta(x_{1}^{0}-x_{2}^{0})\langle0|\big[\phi(x_{1}),\partial_{\nu}^{x_{2}}
\phi(x_{2})\big]|0\rangle = i\hspace{0.02cm}\eta_{\mu 0}\eta_{\nu 0}\delta^{(4)}(x_{1}-x_{2}) \ .
\end{equation}
Collecting terms, we can express the contraction of two gradients of
the $\phi$-field as
\begin{eqnarray} \label{eq:gradprop}
\langle0|\mathcal{T}\big{\{}\partial_{\mu}^{x_{1}}\phi(x_{1})\partial_{\nu}^{x_{2}}
\phi(x_{2})\big{\}}|0\rangle 
& = & i\hspace{0.02cm}\partial_{\mu}^{x_{1}}\partial_{\nu}^{x_{2}}\Delta_{F}^{\phi}(x_{1}-x_{2})-i\hspace{0.02cm}\eta_{\mu 0}\eta_{\nu 0}\delta^{(4)}(x_{1}-x_{2}) \ .
\end{eqnarray}
In the perturbative series of the full propagator for the $S$-field,
the first diagram in Eq.\ (\ref{eq:Pigraph}) also appears in the combination with
two $\phi$-field propagators:
\begin{equation}
\frac{(-ig)^{2}}{2!} \cdot 2\cdot 2
\int\text{d}x^{\prime}\int\text{d}x^{\prime\prime} \ i\Delta(x_{1}-x^{\prime})\,
i\Delta(x^{\prime\prime}-x_{2}) 
\langle0|\mathcal{T}\big{\{}\partial_{\mu}^{x^{\prime}}\phi(x^{\prime})
\partial_{\nu}^{x^{\prime\prime}}\phi(x^{\prime\prime})\big{\}}|0\rangle\langle0|
\mathcal{T}\big{\{}\partial^{\mu,x^{\prime}}\phi(x^{\prime})\partial^{\nu,x^{\prime\prime}}
\phi(x^{\prime\prime})\big{\}}|0\rangle \,.
\end{equation}
Two factors of $-ig$ originate from the three-point vertices in $\hat{\cal L}_{\text{int}}$. 
The factor of $1/2!$ arises because the diagram is second order in perturbation theory. 
A factor of two arises because each $S$ propagator can be joined
with either one of the $S$ legs at the vertex.
Finally, another factor of two comes from the fact that the two $\phi$ lines
at one vertex can be joined with corresponding lines at the other vertex in two
different ways. Successively inserting Eq.\ (\ref{eq:gradprop}) we compute
\begin{eqnarray}
\lefteqn{\hspace*{-1.5cm} 
2g^{2}\int\text{d}x^{\prime}\int\text{d}x^{\prime\prime} \ \Delta(x_{1}-x^{\prime})
\Delta(x^{\prime\prime}-x_{2}) 
\left[i\hspace{0.02cm}\partial_{\mu}^{x^{\prime}}\partial_{\nu}^{x^{\prime\prime}}
\Delta_{F}^{\phi}(x^{\prime}-x^{\prime\prime})-i\eta_{\mu 0}\eta_{\nu 0}\delta^{(4)}(x^{\prime}
-x^{\prime\prime})\right]\langle0|\mathcal{T}\big{\{}\partial^{\mu,x^{\prime}}\phi(x^{\prime})
\partial^{\nu,x^{\prime\prime}}\phi(x^{\prime\prime})\big{\}}|0\rangle}\nonumber \\[0.13cm]
& = & 2g^{2}\int\text{d}x^{\prime}\int\text{d}x^{\prime\prime} \ \Delta(x_{1}-x^{\prime})
\Delta(x^{\prime\prime}-x_{2}) \ i\hspace{0.02cm}\partial_{\mu}^{x^{\prime}}
\partial_{\nu}^{x^{\prime\prime}}\Delta_{F}^{\phi}(x^{\prime}-x^{\prime\prime})
\langle0|\mathcal{T}\big{\{}\partial^{\mu,x^{\prime}}\phi(x^{\prime})
\partial^{\nu,x^{\prime\prime}}\phi(x^{\prime\prime})\big{\}}|0\rangle \nonumber \\
& & - \ i\hspace{0.02cm}2g^{2}\int\text{d}x^{\prime} \ \Delta(x_{1}-x^{\prime})\Delta(x^{\prime}
-x_{2})\langle0|\mathcal{T}\big{\{}\partial^{0,x^{\prime}}\phi(x^{\prime})\partial^{0,x^{\prime}}
\phi(x^{\prime})\big{\}}|0\rangle \nonumber \\
\nonumber \\
& = & -2g^{2}\int\text{d}x^{\prime}\int\text{d}x^{\prime\prime} \ \Delta(x_{1}-x^{\prime})
\Delta(x^{\prime\prime}-x_{2}) \ \partial_{\mu}^{x^{\prime}}\partial_{\nu}^{x^{\prime\prime}}
\Delta_{F}^{\phi}
(x^{\prime}-x^{\prime\prime})\partial^{\mu,x^{\prime}}\partial^{\nu,x^{\prime\prime}}
\Delta_{F}^{\phi}(x^{\prime}-x^{\prime\prime}) \nonumber \\
& & - \ i\hspace{0.02cm}2g^{2}\int\text{d}x^{\prime}\int\text{d}x^{\prime\prime} \ 
\Delta(x_{1}-x^{\prime})\Delta(x^{\prime\prime}-x_{2})\hspace{0.02cm}i\hspace{0.02cm}
\partial_{0}^{x^{\prime}}\partial_{0}^{x^{\prime\prime}}\Delta_{F}^{\phi}(x^{\prime}
-x^{\prime\prime})\delta^{(4)}(x^{\prime}-x^{\prime\prime}) \nonumber \\
& & - \ i\hspace{0.02cm}2g^{2}\int\text{d}x^{\prime} \ \Delta(x_{1}-x^{\prime})
\Delta(x^{\prime}-x_{2})\langle0|\mathcal{T}\big{\{}\partial^{0,x^{\prime}}\phi(x^{\prime})
\partial^{0,x^{\prime}}\phi(x^{\prime})\big{\}}|0\rangle \ .
\end{eqnarray}
With Eq.\ (\ref{eq:Pi12}) one realizes that the last two terms are identical. The final
result is
\begin{align}
-2g^{2}\int\text{d}x^{\prime}\int\text{d}x^{\prime\prime} \ \Delta(x_{1}-x^{\prime})
\Delta(x^{\prime\prime}-x_{2})\partial_{\mu}^{x^{\prime}}\partial_{\nu}^{x^{\prime\prime}}
\Delta_{F}^{\phi}(x^{\prime}-x^{\prime\prime})\partial^{\mu,x^{\prime}}
\partial^{\nu,x^{\prime\prime}}\Delta_{F}^{\phi}(x^{\prime}-x^{\prime\prime}) \nonumber \\
- \ i\hspace{0.02cm}2g^{2}\cdot 2\int\text{d}x^{\prime} \ \Delta(x_{1}-x^{\prime})
\Delta(x^{\prime}-x_{2})\langle0|\mathcal{T}\big{\{}\partial^{0,x^{\prime}}\phi(x^{\prime})
\partial^{0,x^{\prime}}\phi(x^{\prime})\big{\}}|0\rangle \ ,
\end{align}
\end{widetext}
which can be graphically depicted as
\begin{equation} \label{eq:Piregular}
\includegraphics[scale=0.48]{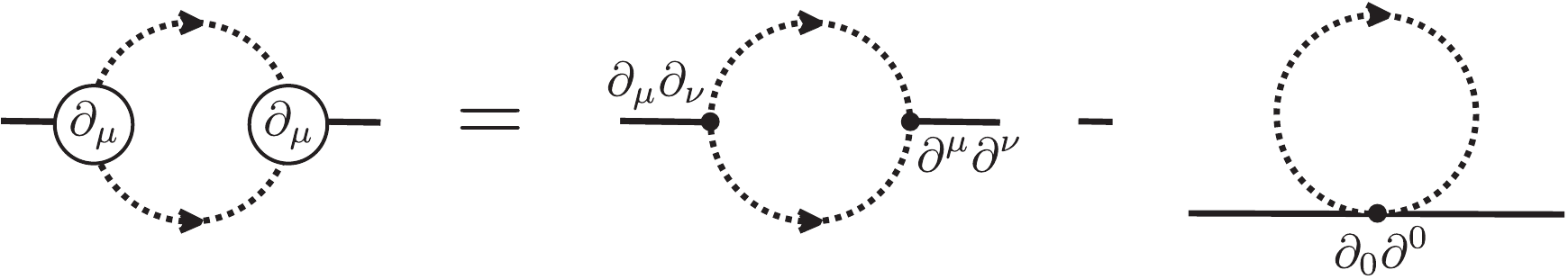} 
\end{equation}
Obviously, the second diagram cancels the tadpole contribution, Eq.\ (\ref{eq:Pi12}), 
in the one-loop self-energy from Eq. (\ref{eq:Pigraph}).

In summary, a derivative interaction in ${\cal L}_{\text{int}}$ produces an additional 
term in the interaction Hamiltonian and thus, after quantization, an additional vertex which
has to be taken into account in perturbative calculations via Feynman rules. In the
one-loop self-energy, this vertex leads to a tadpole diagram. Nevertheless, 
carefully computing contractions between gradients of the field operators we
demonstrated that these lead to a term which exactly cancels the tadpole diagram.
The remaining contribution is exactly equal to the self-energy when computed
with standard Feynman rules using $\hat{\cal L}_{\text{int}}$ and
derivatives acting on the usual Feynman propagators. 

We did not deliver a rigorous proof of this tadpole cancellation to all orders in 
perturbation theory. However, since this seems to be just a demonstration of the validity 
of Matthews's theorem \cite{matthews} which was investigated \emph{e.g.} in 
Refs.\ \cite{duncan,barua,simon,knetter}, we also expect a similar cancellation to work
at higher orders in perturbation theory.

\subsection*{One-loop self-energy from a dispersion relation}

The second way to compute the self-energy is via the dispersion relation (\ref{eq:disp}).
To this end, one needs the imaginary part of the self-energy in order to compute
the real part. The imaginary part can be inferred from the decay width through the
optical theorem. For the one-loop self-energy, the cutting rules imply that the decay width
needs only to be known at tree-level:
\begin{widetext}
\begin{equation} \label{eq:impart}
\text{d}\Gamma \hspace*{2.7cm}
= \ \ -\operatorname{Im}\bigg( \hspace{1.35cm} \bigg) \ 
= \ \ -\operatorname{Im}\bigg( \hspace{1.4cm} \bigg)
\hspace*{-10.5cm}\begin{minipage}{10cm}
\includegraphics[scale=0.31]{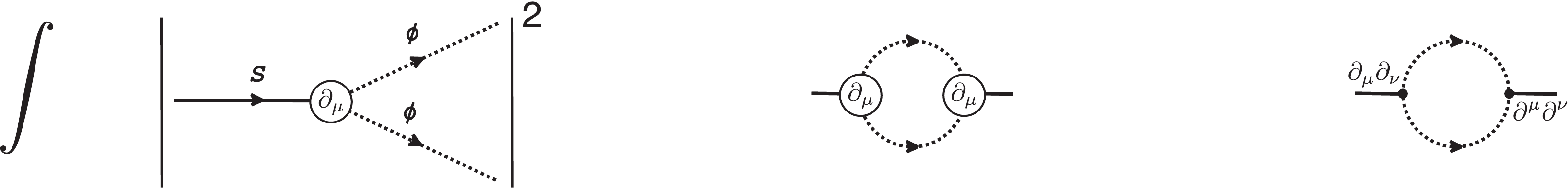} 
\end{minipage} \ \ \ \ \ \ \ .
\end{equation}
The second equality arises from Eq.\ (\ref{eq:Piregular}) and the fact that the tadpole has no 
imaginary part.

The calculation of the tree-level decay width in momentum space
proceeds by replacing derivatives $\partial_{\mu}\rightarrow\pm ip_{\mu}$ (the lower/upper
sign stands for incoming/outgoing particles)
in the Lagrangian (\ref{eq:Lint}), \emph{i.e.}, in our simple model the decay amplitude reads
\begin{equation}
2ig\underbrace{\left(-\frac{s-2m^{2}}{2}\right)}_{=-p_{1}\cdot p_{2}} \ = \ \
\begin{minipage}{7cm}
\includegraphics[scale=0.39]{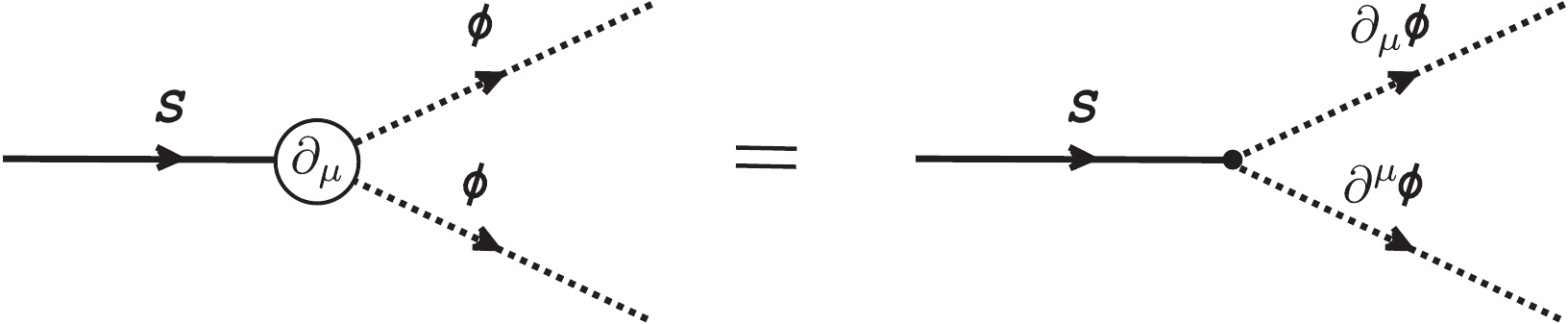}
\end{minipage} \ .
\end{equation}
\end{widetext}
The factor two arises from the two identical particles in the outgoing channel.
The blob in the left diagram represents the vertex as given by Eq.\ (\ref{eq:Hint}), 
while in the middle diagram the above replacement was performed in order to 
calculate the expression on the right-hand side. 
Note that the factor $(s-2m^2)/2$ appears in a similar form in Eq.\ (\ref{eq:eLSMamp}) 
with coupling constant $g=B_{i}^{\text{eLSM}}$. 
Nevertheless, since $s = M^2$, the tree-level decay width is simply a constant. 

Returning now to the imaginary part (\ref{eq:impart}) of the self-energy, we 
observe that, on account of the fact that the tadpole does not contribute
to the imaginary part, with the dispersion
relation (\ref{eq:disp}) one actually only computes the first diagram in Eq.\ (\ref{eq:Pigraph}),
but misses the tadpole contribution. In other words, as we have demonstrated
above, the
first diagram in Eq.\ (\ref{eq:Pigraph}) contains precisely the tadpole contribution, but
with opposite sign, cf.\ Eq.\ (\ref{eq:Piregular}). Consequently, we need to add
this tadpole to the (real part of the) self-energy as computed via the dispersion relation,
in order to have the latter agree with the result obtained from the perturbative calculation.
We remark in passing that the emergence of a tadpole can also be explicitly 
demonstrated by cleverly
manipulating the expression for $\Pi(s)$ as it results from the Feynman rules. 
In the case of our effective model from Sec.\ \ref{sec:section3} we computed the
self-energies precisely in the manner explained above, \emph{i.e.}, from a
dispersion relation and adding the corresponding tadpole diagrams. 

We conclude by remarking that, in the eLSM Lagrangian for the scalar--isovector state, 
derivatives also occur in front of the decaying field $a_{0}$; these are the terms with 
coupling constants $C_{i}^{\text{eLSM}}$ in Eq.\ (\ref{eq:Lagrangian}).
All that has been stated above applies also in this case, with the exception that, besides constant tadpole terms, 
also $s$-dependent contributions appear in the expression for $\Pi(s)$. We note that all these additional contributions 
also cancel in a similar way as we discussed above.

\bibliography{a0revisited_08dec}

\end{document}